\documentclass[10pt,conference,letterpaper]{IEEEtran}

\IEEEoverridecommandlockouts
\usepackage{cite}
\usepackage{amsmath,amssymb,amsfonts}
\usepackage{algorithmic}
\usepackage{graphicx}
\usepackage{textcomp}
\usepackage{xcolor}
\usepackage{tablefootnote}
\usepackage{siunitx}
\usepackage{makecell}
\usepackage{booktabs}
\usepackage{todonotes}
\usepackage{algorithm}
\usepackage{url}
\usepackage[acronym]{glossaries}
\usepackage{blindtext}

\makeglossaries

\def\BibTeX{{\rm B\kern-.05em{\sc i\kern-.025em b}\kern-.08em
    T\kern-.1667em\lower.7ex\hbox{E}\kern-.125emX}}
\begin{document}

%\newacronym{dm}{DM}{Decision Making}
\newacronym{pso}{PSO}{Particle Swarm Optimization}
\newacronym{sdv}{SDV}{Software Defined Vehicles}

%\newacronym{cv}{CV}{Connected Vehicles}

\newacronym{oem}{OEM}{Original Equipment Manufacturer}
\newacronym{qos}{QoS}{Quality of Service}
\newacronym{rtt}{RTT}{Round Trip Time}
\newacronym{iot}{IoT}{Internet of Things}
\newacronym{iov}{IoV}{Internet of Vehicles}
\newacronym{mec}{MEC}{Mobile Edge Computing}
\newacronym{v2v}{V2V}{Vehicle-to-Vehicle}
\newacronym{eta}{ETA}{Estimated Time of Arrival}
\newacronym{rsu}{RSU}{Road-Side Units}
\newacronym{dnn}{DNN}{Deep Neural Networks}
\newacronym{api}{API}{Application Programming Interfaces}
\newacronym{p2p}{P2P}{peer-to-peer}
\newacronym{dm}{DM}{decision-making}
\newacronym{rl}{RL}{reinforcement learning}
\newacronym{cicd}{CI/CD}{Continuous Integration/Continuous Deployment}
\newacronym{ip}{IP}{Internet Protocol}
\newacronym{hvac}{HVAC}{Heating Ventilation and Air Conditioning}
\newacronym{bf}{BF}{Brute-Force}

\newacronym{cpu}{CPU}{Central Processing Unit}
\newacronym{gpu}{GPU}{Graphics Processing Unit}
\newacronym{ml}{ML}{Machine Learning}

% conference idea (6 Seiten + 150$ pro Seite):
% https://www.cloud-conf.net/cscloud/2025/ssc/submission.html
%\title{A Scalable Multilayer Pipeline for Computation Offloading in Software-Defined Vehicles}
\title{Towards Intelligent Computation Offloading in Dynamic Vehicular Networks: A Scalable Multilayer Pipeline}

\makeatletter
\newcommand{\linebreakand}{%
  \end{@IEEEauthorhalign}
  \hfill\mbox{}\par
  \mbox{}\hfill\begin{@IEEEauthorhalign}
}
\makeatother

\author{
\IEEEauthorblockN{Falk Dettinger, Matthias Weiß, Baran Can Gül, Sruthi Mangala Suresh, Nasser Jazdi and Michael Weyrich}
\IEEEauthorblockA{
\textit{Institute of Industrial Automation and Software Engineering (IAS)} \\
\textit{University of Stuttgart} \\
Pfaffenwaldring 47, 70550 Stuttgart, Germany \\
E-Mail: \{falk.dettinger, matthias.weiss, baran-can.guel, sruthi.suresh, nasser.jazdi, michael.weyrich\}@ias.uni-stuttgart.de}}

\maketitle
\begin{abstract}
\glsentrylong{sdv} face an increasing computational gap as advanced algorithms and frequent software updates demand more processing power while onboard hardware remains static throughout a vehicle's 10+ year lifespan. This mismatch threatens the performance of safety-critical functions including advanced driver-assistance systems and real-time perception tasks. We propose a novel four-layer computation offloading pipeline that dynamically distributes vehicular functions to cloud and edge resources while meeting strict \glsentrylong{rtt} constraints. Our key contribution is an enhanced \glsentrylong{pso} algorithm that integrates distance- and direction-based penalties with functional requirements to optimize edge server selection for mobile vehicles. Evaluation using a Kubernetes-based cloud infrastructure with realistic vehicular mobility patterns demonstrates that our approach reduces average response time compared to conventional \glsentrylong{bf} methods while maintaining the success rate for latency-critical tasks. The modified \glsentrylong{pso} algorithm achieves an average execution time of 26\,ms across ten servers and tasks on \glsentrylong{cpu}, and 550\,ms across 15 servers with 1000 tasks on \glsentrylong{gpu}. These results confirm the pipeline's effectiveness in bridging the computational gap for next-generation \gls{sdv}.
\end{abstract}


\section{Introduction}\label{sec:introduction}
The automotive industry is shifting toward \gls{sdv}, where functions increasingly rely on advanced algorithms rather than mechanical systems \cite{baumann2024total}. These vehicles run hundreds of control units for autonomous driving, infotainment, and predictive maintenance, requiring frequent updates and growing computational power to meet safety and performance standards \cite{Guel24_2}.

A critical mismatch exists between growing software demands and static vehicle hardware, which remains unchanged over a typical lifespan exceeding a decade \cite{KBA2024, SPGlobal2024}. This widening gap risks computational obsolescence, potentially compromising safety-critical functions such as collision avoidance and real-time traffic analysis \cite{Mizrachi2025}.

Computation offloading offers a viable solution by leveraging cloud and edge resources to execute computationally intensive vehicular functions. However, vehicles operate under strict real-time constraints \cite{Stuempfle2025SDV, praveen2025autonomous} and rely on a heterogeneous set of functions within rapidly changing server accessibility and signal quality \cite{fi16040108, weiss2025sdvdiag}.  In SDVs, offloading typically shifts demanding perception or decision‑making steps to edge or cloud servers when onboard hardware reaches its limits, while lightweight local models remain on the vehicle \cite{10.1007/978-3-031-62277-9_38}. Vehicles may also retrieve updated models from backend servers, which are then executed locally or offloaded depending on their computational demands. This tight coupling of local and backend capabilities requires an offloading strategy that adapts to connectivity, server availability, and function‑specific timing constraints. 

Existing offloading approaches often target static mobile devices or isolated components, overlooking the unique demands of software-defined vehicles. Most rely on simplistic server selection that ignores mobility patterns, directional movement, and diverse latency requirements. 

We propose a four‑layer offloading framework for SDVs, comprising extraction, decision, execution, and detection stages. Rather than relying on a single optimization method, the framework supports diverse decision‑making algorithms for selecting suitable edge or cloud resources under dynamic connectivity conditions. It anticipates future communication constraints by incorporating spatial, temporal, and functional context into the offloading process. We evaluate the framework on a Kubernetes‑based cloud infrastructure using realistic mobility simulations, server disturbances, and latency‑sensitive workloads.

The remainder of this paper is organized as follows. Section~\ref{sec:background} introduces fundamental concepts and challenges in vehicular computation offloading. Section~\ref{sec:related_work} reviews existing approaches and identifies research gaps. Section~\ref{sec:concept} details our four-layer pipeline architecture and enhanced PSO algorithm. Section~\ref{sec:evaluation} presents comprehensive experimental results, and Section~\ref{sec:conclusion} concludes the paper.

\section{Background}\label{sec:background}
To conserve computational resources and energy on mobile devices such as smartphones, IoT systems, and connected \gls{sdv}s, computation offloading delegates intensive tasks to high-performance backend servers \cite{acheampong2023parallel}. Offloading occurs in two forms: binary (entire task) and partial (selected components).

%The computation offloading process is addressed in various studies, each offering a distinct framework while emphasizing shared principles. Feng et al. \cite{feng2022computation} describe the process as a series of steps that involves dividing tasks into subtasks, making decisions about whether and where to offload, deploying the tasks to selected servers, processing these tasks, and returning the results to the user. Nguyen et al. \cite{nguyen2020smartphone} focus on phases involving the analysis of resource usage, the partitioning of applications, and decision-making based on runtime profiling. Similarly, Zhang et al. \cite{zhang2024survey} propose an approach that begins with task generation, proceeds through resource and task allocation (guided by optimization objectives such as energy efficiency and latency), continues with data communication and processing, and culminates in final execution decisions. While these frameworks differ in structure, they all underline essential elements such as task analysis and partitioning, decision-making, the offloading and deployment, and execution of services, which serve as the backbone of computation offloading systems.
Multiple frameworks describe the offloading process, typically involving task partitioning, decision-making, deployment, and execution \cite{feng2022computation, nguyen2020smartphone, zhang2024survey}. While structural details vary, core elements remain consistent across domains.

%These core components depend on a robust backend infrastructure to operate effectively. Cloud and edge servers provide this necessary support \cite{10757637}. Although centralized cloud servers offer nearly unlimited scalability and performance for handling resource-intensive tasks, their physical distance from clients limits their ability to deliver real-time results. In contrast, decentralized edge servers, situated closer to clients, enable real-time processing, albeit with reduced scalability and overall performance. Integrating either offloading approach in a productive scenario requires careful monitoring of the involved metrics. The offloading process is then commonly triggered by performing anomaly detection on the measured time series, for which the thresholds may be given by the performance requirements of the applications or the offloading process itself \cite{weiss2024ad, weiss2024simulating}.
These core components depend on a robust backend infrastructure to operate effectively. Cloud and edge servers provide this necessary support \cite{10757637}. In this context, centralized cloud servers offer scalability but suffer from latency due to physical distance while Edge servers enable real-time processing closer to clients, though with limited capacity. Integrating offloading approaches in a productive scenario requires careful monitoring of the involved metrics. The offloading process is then commonly triggered by performing anomaly detection on the measured time series, for which the thresholds may be given by the performance requirements of the applications or the offloading process itself \cite{weiss2024ad}.%, weiss2024simulating}.

%\todo[inline]{folgendes evtl. ausklammern}
%\begin{itemize}
%    \item notwendigkeit der Cluster (Kubernetes) und Ausdehnung des Clusters auf die Edge -- wegen er notwendigen Sklaierbarkeit ist es nur schwer möglich die Skalierbarkeit ohne Cluster zu realisieren.
%    \item Deployment loop
%    \item Feedback loop
%\end{itemize}

% New related work: 19.08.2025

\section{Related Work}\label{sec:related_work}
Computation offloading for vehicular networks has shifted decisively toward learning-based and hybrid decision engines~\cite{choudhury2024machine}. Farimani et al.~\cite{farimani2024deadline} apply deep reinforcement learning with explicit deadline constraints in vehicular edge networks, while %Dai et al.~\cite{dai2024meta} propose meta-reinforcement learning to improve generalization across heterogeneous multi-task offloading scenarios. 
Wei et al.~\cite{wei2024many} address many-to-many offloading through multi-agent reinforcement learning% in vehicular fog computing
. Although these methods achieve strong algorithmic performance, they optimize isolated decision engines without addressing end-to-end pipeline integration or deployment on physical infrastructure.
 
Mobility-aware strategies form a second research thread. Li et al.~\cite{li2024mobility} jointly model mobility patterns and task dependencies for intelligent vehicles, and Xia et al.~\cite{xia2023location} propose location-aware delay minimization in \gls{mec}. Ling et al.~\cite{ling2024qos} dynamically select offloading locations based on estimated arrival times, optimizing \gls{qos} fairness among \glspl{rsu}. %Zhao et al.~\cite{zhao2024adaptive} combine swarm intelligence with digital twins for adaptive offloading in vehicular edge computing. 
However, these approaches typically consider position or trajectory in isolation and do not jointly incorporate directional movement relative to servers, distance-based penalties, and function-specific \glsentrylong{rtt} thresholds into the offloading decision.
 
At the system level, end-to-end offloading frameworks remain rare. Rehman et al.~\cite{rehman2023foggyedge} propose FoggyEdge, an information-centric fog-edge architecture for vehicular computing. While FoggyEdge addresses data routing and caching at the naming layer, it does not incorporate mobility-dependent spatial penalties, per-function latency requirements, or feedback-driven adaptation. Zhang et al.~\cite{zhang2024computation} explore trajectory-based pre-offloading, and Xu et al.~\cite{xu2025digital} present a digital-twin-assisted blockchain-based vehicular edge architecture. Despite these contributions, the vast majority of existing frameworks are validated exclusively in simulation and lack mechanisms for online feedback and adaptive decision refinement~\cite{choudhury2024machine}.
 
Our work addresses these gaps through a deployed four-layer pipeline on Kubernetes infrastructure. Unlike existing approaches, the system integrates direction- and distance-based spatial penalties with function-specific \glsentrylong{rtt} constraints directly into the decision logic, and includes a Detection Layer for feedback-driven adaptation. This combination of real-infrastructure deployment, mobility-aware spatial penalties, and closed-loop adaptation distinguishes our pipeline from prior simulation-only studies.

\section{Dynamic Decision-Making Framework for Computation Offloading}\label{sec:concept}
%As described in Section \ref{sec:background}, conventional computation offloading pipelines typically follow four steps: task analysis and partitioning, decision-making, offloading and deployment, and service execution. Building on this foundation, our research proposes a broader four-stage structure according 
Building on conventional offloading pipelines (see Section \ref{sec:background}), we propose a four‑layer architecture (Fig. \ref{fig:plot_decision_edge}). The first stage is the \textbf{Extraction Layer}, responsible for extracting relevant task information received from vehicles and collecting server data from computation clusters. The subsequent \textbf{Decision Layer} determines the optimal offloading strategy based on the extracted information. Once a decision is made, it is communicated to the vehicle through the \textbf{Offloading and Execution Layer}, enabling the function to be executed either locally or in the backend. %This implicitly assumes that the backend is aware of all available functions and caching is not a problem. 
Because the decision-making process relies heavily on predicted system states, we have incorporated a \textbf{Detection Layer} as a feedback loop to dynamically adjust decisions in response to changing conditions. The following subsections detail each layer further.

\begin{figure}[tb]
    \centering
    \includegraphics[width=\columnwidth]{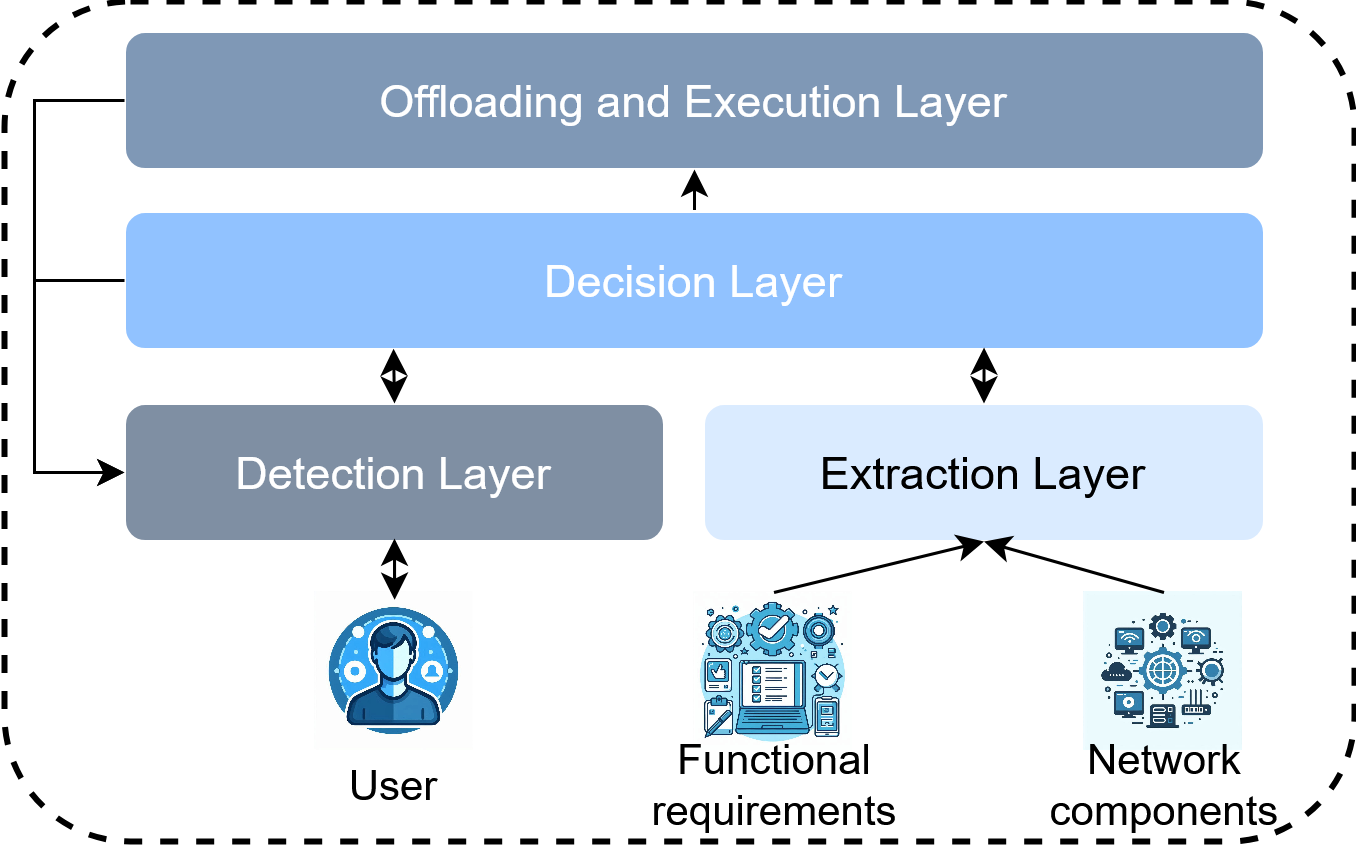}
    \caption{Architectural overview of a four step pipeline structure for computation offloading considering the Layers \textit{Extraction}, \textit{Decision}, \textit{Offloading and Execution}, and \textit{Detection Level}.}
    \label{fig:plot_decision_edge}
\end{figure}

\subsection{Extraction Layer}\label{sec:extraction}

The Extraction Layer is responsible for extracting information about functions and available backend servers, organizing them in a \textbf{Service List} and a \textbf{Server List}. Both sets of information are directly accessible to the decision layer for proper assignment of functions to servers. The \textbf{Service List} is a file in which functions are evaluated based on specific requirements, following the criteria outlined by Sommer et al. \cite{10.1007/978-3-031-62277-9_38}. Functions are evaluated based on requirements such as criticality, timing constraints, and resource demands. Experts are assumed to define these requirements during the development phase of a function, allowing them to be directly considered when requesting computation offloading.

%The \textbf{Server List} contains information on available backend servers, their current state, and metadata necessary for precise decision-making. There are two possible ways a server can be integrated into the offloading pipeline: (i) as part of a computation cluster or (ii) as a standalone server. In case (i), the cluster's \gls{api} allocates the server information and transfers it to the decision-making server. In case (ii), a direct \gls{p2p} connection between the decision-making server and the backend server must be established, with the backend server sending heartbeat signals containing its current system state.
The \textbf{Server List} contains all relevant information about available backend servers, including their current state and metadata required for decision‑making. Server information is obtained either through the computation cluster, which exposes the necessary data via its API, or through direct communication with standalone servers. In both cases, the goal is to maintain an up‑to‑date view of server availability and load so that the decision layer can reliably evaluate offloading options.

\subsection{Decision Layer}\label{sec:decision}

The assignment of tasks to backend servers takes place in the \textbf{Decision Layer}, which operates in two stages. Decisions are based on assumptions about the overall system state, including, but not limited to the predicted transmission time (\glsentrylong{rtt}), computation time, and system utilization of backend servers. These metrics influence one another and vary depending on the selected server and service combination.

%In the first stage, the expected system behavior is predicted for each offloading option. Specifically, a CNN-LSTM model is used to estimate transmission time, while separate LSTM and CNN-LSTM models are employed to predict computation time and system utilization, respectively. These models are trained on historical offloading data and contextual features, enabling robust predictions under dynamic conditions. The resulting estimates form the basis for evaluating the feasibility and efficiency of each server-service pair. 

%In the second stage, the actual offloading decision is made using reinforcement learning or meta-heuristic methods. These algorithms process the predicted metrics alongside contextual penalties such as directional alignment, distance to server, and remaining stay time within communication range. For example, a directional penalty is calculated using cosine similarity according to equation \ref{eq:cosine_similarity} between the vehicle’s movement vector and the vector pointing to the server. The penalty factor, defined in Equation~\eqref{eq:direction_penalty}, is added to the communication time, being zero when the vehicle is heading toward the server and two when heading away. A distance penalty further prioritizes closer servers, and incorporating the remaining stay time helps avoid offloading to servers that will soon be out of range. These considerations are particularly relevant for edge computing, where low-latency, real-time processing is required.
In the first stage, system behavior is predicted using trained models: a CNN-LSTM estimates transmission time, while ML models predict transmission time, computation time, and utilization. These models use historical offloading data and contextual features to enable robust predictions under dynamic conditions.

In the second stage, reinforcement learning or meta-heuristics select the optimal server, incorporating penalties for directional alignment, distance, and remaining stay time. Directional penalties (Eq.~\ref{eq:direction_penalty}) are computed via cosine similarity between the vehicle’s movement vector and the server vector, and added as a weighted offset to the communication time. These factors are essential because they prevent offloading to servers the vehicle will soon move out of range of, which could lead to rising RTTs or even complete loss of the result when the connection drops.

\begin{equation}
\text{DirectionPenalty} 
  = 1 + \frac{\mathbf{A} \cdot \mathbf{B}}
              {\|\mathbf{A}\| \, \|\mathbf{B}\|},
\label{eq:direction_penalty}
\end{equation}

\noindent where 
$\mathbf{A} = (dir_{v_x}, dir_{v_y})$ is the vehicle movement direction, 
and $\mathbf{B} = (dir_{s_x}, dir_{s_y})$ is the direction to the server. 
Here, $\mathbf{A} \cdot \mathbf{B} = \sum_{i=1}^{n} A_i B_i$, 
$\|\mathbf{A}\| = \sqrt{\sum_{i=1}^{n} A_i^2}$, 
and $\|\mathbf{B}\| = \sqrt{\sum_{i=1}^{n} B_i^2}$.

%\begin{equation}
%\text{DirectionPenalty} = 1 - CosineSimilarity
%\label{eq:direction_penalty}
%\end{equation}

%\begin{equation}
%penalty(t) = \,
%\label{eq:stay_time}
%\end{equation}
\begin{figure*}[tb]
    \centering
    \includegraphics[width=0.85\linewidth]{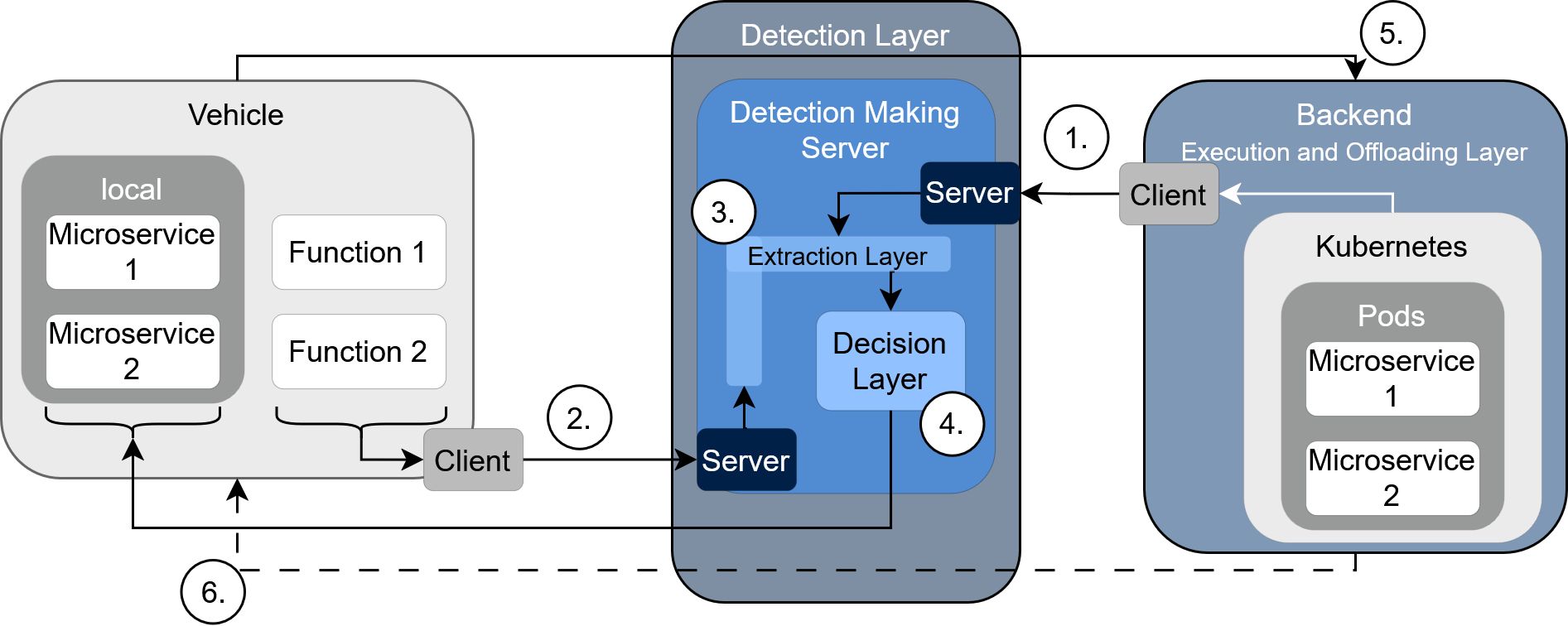}
    \caption{Proposed Offloading Architecture for vehicular functions. A centralized decision-making server is used for extracting relevant information and decision-making, while the vehicle is able to execute a function locally or in the backend. The backend contains cloud and edge server orchestrated by a computation cluster. The numbers indicate the process sequence, starting with 1 and finishing with 6.}
    \label{fig:pipeline_overview}
\end{figure*}

The \glsentrylong{dm} process operates as follows. The algorithm first receives all relevant inputs from the Extraction Layer, including the requested functions with their requirements, the vehicle’s current position, and server‑side information such as location, communication range, and utilization. A direction‑dependent penalty is added to the communication‑time estimates to account for vehicle movement. Based on these inputs, the algorithm checks whether each server is within range and whether the remaining stay time is sufficient for completing the service. Servers that fail these feasibility checks are marked as invalid. In parallel, the Prediction Layer provides forecasts of expected utilization, which are incorporated as additional constraints. Using this combined information, the algorithm selects the optimal offloading target and returns the server’s IP address and port to the vehicle, which then initiates the actual offloading request via the Offloading and Execution Layer.

\subsection{Offloading and Execution Layer}\label{sec:offloading}
Computation offloading requires scalable, low-latency function execution, as multiple vehicles may request the same function simultaneously. To address this, we employ a computation cluster for orchestrating servers, managing function caching, and scaling deployments across clients with minimal delay. This ensures flexibility even under dynamic routing conditions. A \gls{cicd} loop further enhances the cluster by enabling rapid function updates.

\subsection{Detection Layer}\label{sec:detection}
In a dynamic network, servers can unexpectedly fail, be shut down, or be updated. Vehicles enter and exit the communication range of the backend servers. To respond to these constant changes, the \textbf{Detection Layer} employs a feedback mechanism to continuously adapt the offloading pipeline to changing conditions.

The process consists of two steps. In step (i), previous offloading decisions are evaluated by comparing predicted metrics (e.g., RTT and utilization) with actual measurements. Decisions that meet the requirements are marked as "successful", otherwise, priority‑based penalties are applied until they are classified as "incorrect". Thresholds depend on functional criticality, with non‑critical services tolerating delays and safety‑critical tasks requiring strict limits.
In step (ii), classification results are used to train supervised learning models that refine the prediction and decision-making stages. Thereby considering input features including predicted RTT, system utilization, spatial penalties, and functional criticality. 

\subsection{Pipeline Summary}\label{sec:summary}
%Fig. \ref{fig:pipeline_overview} illustrates the overall system architecture. Depending on system complexity, multiple equally autonomous offloading servers with independent domains can coexist. Furthermore, the computing hardware is organized into clusters that combine both cloud and edge servers, a design that simplifies the deployment of services while significantly enhancing system responsiveness and scalability. 
Fig.~\ref{fig:pipeline_overview} illustrates the system architecture. Depending on complexity, multiple autonomous offloading servers may operate in parallel across independent domains. Computing hardware is clustered across cloud and edge resources to streamline service deployment and improve responsiveness.

%The process begins (1) with server state updates from the cluster to the \glsentrylong{dm} server. Vehicles transmit offloading requests and functional requirements (2), prompting updates to the function and server lists (3). Based on this, the offloading decision is made (4), and the selected \gls{ip} and port are returned to the vehicle (5), which initiates execution. Results are then transmitted back to the vehicle (6).
The process begins (1) with server state updates from the cluster to the \glsentrylong{dm} server. Vehicles transmit offloading requests and functional requirements (2), which update the function list on the decision‑making server, while the server list is refreshed using the latest cluster information (3). Based on this combined state, the offloading decision is made (4), and the selected \gls{ip} and port are returned to the vehicle (5), which initiates execution. Results are then transmitted back to the vehicle (6).

\section{Evaluation}\label{sec:evaluation}

For the evaluation, we define three test cases to assess the performance of the proposed offloading pipeline. The cases are structured to reflect a progression from decision quality to system robustness and computational efficiency:

\begin{enumerate}
    \item \textbf{Impact on Decision-Making:} Impact on Decision-Making: We evaluate the correctness and sensitivity of the decision logic using predefined functions and \gls{rtt} values sampled from empirically bounded distributions \cite{dettinger2025directives}.
    \item \textbf{Decision-Making Time and Accuracy:} We assess computational efficiency and accuracy by comparing brute‑force reference decisions with \gls{pso}‑based optimization.
    \item \textbf{Communication with Cloud/Edge:} We evaluate robustness under dynamic conditions, including backend masking, distance‑based availability, and error handling in the \gls{dm} server.
\end{enumerate}

\iffalse
For the evaluation, we define three test cases to assess the performance of the proposed offloading pipeline. The cases are summarized in the enumerated list below, corresponding to the order of descriptions in Section \ref{sec:result}.

\begin{enumerate}
    \item \textbf{Impact on Decision-Making:} The effects of the direction-based penalty and functional requirements are evaluated using using predefined functions with predicted \glsentrylong{rtt} values sampled from empirically bounded distributions \cite{dettinger2025directives}.
    \item \textbf{Communication with Cloud/Edge:} We assess whether our pipeline can effectively handle different types of backend servers, including cloud and edge environments. This is achieved by masking the backend server positions and limiting service availability based on distance. In this context, we also verify whether the pipeline’s error-handling mechanism can dynamically update the list of available servers for \glsentrylong{dm}.
    \item \textbf{Decision-Making Time and Accuracy:} This evaluation examines the duration of \glsentrylong{dm} and the expected computation time. It is assumed that a task has full access to all available resources on a device at any given time. A randomized set of tasks is used to compute reference best-case and worst-case scenarios via a brute-force approach. These results are then compared to the \gls{pso} \glsentrylong{dm} algorithms. Moreover, the total processing time of the pipeline is measured.
\end{enumerate}
\fi

\subsection{Evaluation Setup}\label{sec:setup}
%During the evaluation, a simulated vehicle—characterized by its position, direction, and velocity—requests computation offloading from a \glsentrylong{dm} server, which is co-located with the vehicle simulation. The offloaded functions consist of object recognition and emotion recognition algorithms~\cite{gul2025fedmultiemo}, both containerized and deployed on the Kubernetes cluster as well as locally on the vehicle.
During evaluation, a simulated vehicle characterized by position, direction, and velocity requests computation offloading from a co-located \glsentrylong{dm} server. The offloaded functions include object and emotion recognition algorithms, containerized and deployed both locally and on a Kubernetes cluster.

\begin{figure*}[t]
    \centering
    \includegraphics[width=0.98\linewidth]{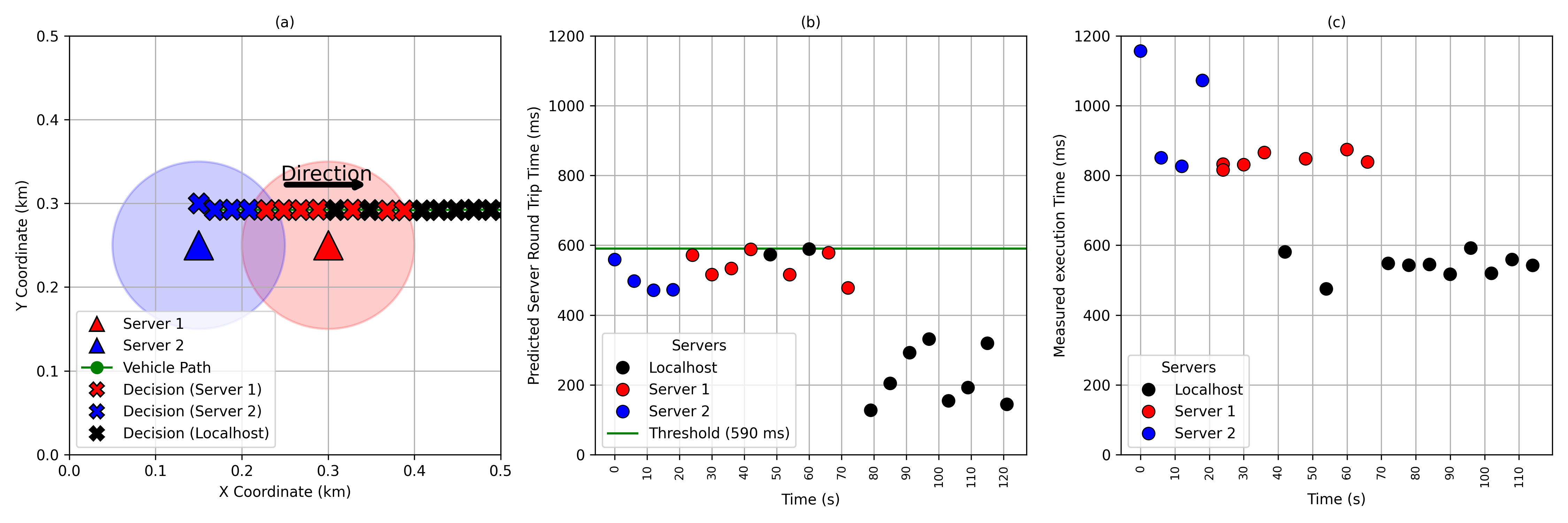}
    \caption{Offloading decision for object recognition based on predicted and measured \gls{rtt}, considering three destinations: local (black), Server 1 (red), and Server 2 (blue). (a) shows the spatial setup with server locations and offloading decisions along a simulated trajectory. (b) visualizes the decision based on the predicted round trip times and the decision threshold (590\,ms). (c) presents measured execution times, revealing significant deviations from predictions.)}
    \label{fig:plot_requirements}
\end{figure*}

Two functions are used for the evaluation. The emotion‑recognition function processes inputs of around 140\,kB per inference and exhibits non‑deterministic RTTs between 500\,ms and 750\,ms depending on system load. The object‑recognition function shows similar latency characteristics (400–750\,ms), making both suitable as realistic test cases for offloading decisions.
%The size of the input data is around 140 kB per inference and exhibits non-deterministic execution times (RTT) ranging from 500\,ms to 750\,ms, depending on system load. The object recognition function is based on a YOLOv8 model using pretrained yolov8n weights~\cite{yolov8_ultralytics}, with slightly smaller execution time ranging from 400\,ms to 750\,ms, but similar variability in execution time. These characteristics inform the functional requirement threshold and serve as realistic test functions for latency-sensitive offloading decisions.

The evaluation uses a cloud‑based Kubernetes cluster consisting of one control and two worker nodes, accessible via the Wide Area Network. This setup provides realistic conditions for dynamic orchestration, as services may migrate between nodes depending on resource availability and network load. To enable controlled scalability experiments, the physical locations of the backend servers are abstracted in software, allowing direction‑ and distance‑based penalties to be evaluated independently of the actual cluster topology. This combination of real infrastructure and simulated spatial dynamics enables reproducible tests under representative edge‑computing constraints.

The \gls{pso} component is not tied to the pipeline design and serves only as an example decision algorithm, while any method consuming the same inputs could be used instead. \gls{pso} is chosen because it handles partially known problem statements and remains applicable in dynamic environments without retraining. In this evaluation, we use a small swarm (10 particles) with standard parameter settings (inertia 0.5, cognitive/social coefficients 1.5) and 80 iterations. The fitness function minimizes the maximum server load, and particle positions are discretized to server IDs.

Vehicle and server positions are simulated, while \gls{rtt} values are sampled from empirically derived distributions based on thousands of cloud executions. To isolate the behavior of the decision‑making logic, both the Prediction Layer and the Detection Layer are disabled during evaluation. Although the architecture supports \gls{ml} models for forecasting \gls{rtt}, computation time, and utilization, we rely on empirical \gls{rtt} distributions to avoid mixing prediction accuracy with decision performance.

The reported results therefore refer exclusively to the evaluated components—namely data extraction, the PSO‑based decision‑making logic, and cloud/edge execution. They do not imply end‑to‑end robustness of the full pipeline. Real‑time correctness under dynamic and volatile network conditions remains outside the scope of this evaluation and will be addressed in future work.

\subsection{Results}\label{sec:result}
In the following, the results of the evaluation cases outlined in Section \ref{sec:setup} are summarized and discussed. The following order matches the order in the enumerated list in Section \ref{sec:setup}

\subsubsection{Impact on Decision-Making}
%Initially, the impact of the proposed \gls{pso} modifications on the \glsentrylong{dm} is evaluated using functional requirements and a direction-based penalty. The evaluation considers two backend servers and local execution, with a fixed \glsentrylong{rtt} threshold of 590\,ms for object recognition and 630\,ms for emotion recognition. Functions are offloaded every 5\,ms, and the actual \glsentrylong{rtt} varies dynamically.
We first evaluate the impact of functional requirements and direction‑based penalties on the decision‑making process. Two backend servers and local execution are considered, with \gls{rtt} thresholds of 590\,ms (object recognition) and 630\,ms (emotion recognition). Offloading decisions are made every 5\,ms while \gls{rtt} values vary dynamically.

Fig. \ref{fig:plot_requirements} summarizes the offloading behavior for the object‑recognition task.
Color coding is consistent across all subplots: local (black), Server 1 (red), Server 2 (blue), while dots and crosses represent the decisions.

Subplot (a) shows the simulated trajectory, the positions of both backend servers (red and blue triangles), their communication ranges (shaded areas), and the resulting offloading decisions (cross markers). Tasks are offloaded to the fastest server that satisfies the functional requirements. If no server is available or all violate the threshold, local execution is selected.

Subplot (b) shows the offloading decisions derived from predicted RTT values, which are sampled from empirically bounded distributions obtained from real cloud measurements. The decision threshold of 590\,ms defines whether a server is considered feasible. As the vehicle moves through the communication ranges of both servers, the predicted RTTs fluctuate accordingly, leading to alternating selections between Server 1 and Server 2 whenever their predicted values fall below the threshold. When both servers either exceed the threshold or are out of range, the decision logic correctly falls back to local execution. This subplot therefore illustrates how the decision‑making behaves under idealized prediction conditions and how predicted RTTs interact with communication range constraints.

Subplot (c) compares the predicted RTTs with measured execution times on the Kubernetes cluster. Measured RTTs deviate by up to 200\,\% from predictions, causing all offloaded tasks to exceed the threshold post‑execution. This demonstrates that prediction errors can lead to functional deadline violations even when the predicted values appear compliant.

%To further evaluate the \glsentrylong{dm} process, the simulated vehicle follows an elliptical path around two edge servers. The computation time is set to a constant value for both servers to isolate evaluation of penalties based on direction and distance. Furthermore, only the object recognition algorithm is considered for \glsentrylong{dm}. The results are shown in Fig.~\ref{fig:plot_elliptic}, where the gray arrows indicate the vehicle’s movement direction.

To further evaluate the \glsentrylong{dm} process, the simulated vehicle follows an elliptical trajectory around two edge servers. Computation times are kept constant to isolate the effect of direction‑ and distance‑based penalties, and only the object‑recognition function is considered. Fig. \ref{fig:plot_elliptic} shows the resulting decisions, with gray arrows indicating the vehicle’s movement direction.

%The green horizontal line illustrates the behavior when only the direction-based penalty is considered. In this case, the server with the highest cosine similarity (see Eq. \ref{eq:direction_penalty}) is selected. Meanwhile, the green dashed vertical line reflects the behavior when only the distance-based penalty is applied, resulting in selection of the nearest server. This setup is designed to isolate the influence of direction- and distance-based penalties. In the upper-right and lower-left quadrants in Fig. \ref{fig:plot_elliptic}, the offloading decision shifts depending on the relative weighting of both penalties: the stronger the emphasis on distance, the more the decision tends toward selecting the closest server. The blue and red crosses represent individual offloading decisions under these conditions.
The green horizontal line illustrates the case where only the direction‑based penalty is applied, selecting the server with the highest cosine similarity (see Eq. \ref{eq:direction_penalty}). The green dashed vertical line shows the behavior when only the distance‑based penalty is used, resulting in selection of the nearest server. This setup isolates the influence of both penalty terms.

When both penalties are active, the offloading decision shifts depending on their relative weighting. In the upper‑right and lower‑left quadrants, a stronger distance weighting moves the decision toward the nearest server, while a stronger direction weighting favors the server aligned with the vehicle’s movement. The red and blue crosses in Fig. \ref{fig:plot_elliptic} represent the resulting offloading decisions under these conditions.

%This partial evaluation demonstrates that, under the assumption of equal computation times, offloading decisions can be adapted based on direction- and distance-based penalties by incorporating vehicle behavior into the decision-making process. In real-world scenarios, however, fluctuations in RTT values may override both penalty components, leading the system to select the fastest server instead.

Under equal computation times, this demonstrates how direction‑ and distance‑based penalties can steer the decision‑making process. In real deployments, however, RTT fluctuations typically dominate both penalties, causing the system to select the fastest server instead.

%ersetzt am 23.10.25 durch obigen Absatz
%The blue and red crosses represent offloading decisions based solely on the direction-based penalty. The green horizontal line indicates the behavior when only the direction-based penalty is considered. In contrast, the green dashed line represents the split, when the distance-based penalty outweighs the direction-based penalty. Then the closest server is selected. This setup is purely selected to show the impact of distance and direction based penalties. In real-world scenarios, RTT fluctuations may override direction/distance penalties, then the fastest server is selected. 

\subsubsection{Decision-Making Time and Accuracy}

To evaluate decision‑making accuracy and performance, the modified PSO algorithm is benchmarked against a brute‑force (BF) reference using randomized functions, assumed RTT values, and varying numbers of servers. Since the evaluated problem sizes allow exact solutions, BF provides the global optimum and thus bounds the behavior of all heuristic methods. The PSO algorithm closely matches the BF best‑case results, with an average deviation of 0.7\,\% for ten servers and tasks.

Runtime measurements show 26\,ms on CPU and 62\,ms on GPU for this configuration. The CPU is initially faster due to GPU initialization and transfer overheads, but after roughly 50 tasks the GPU amortizes these costs and becomes faster. Across up to 15 servers and 1,000 tasks, PSO scales nearly linearly, requiring 2.5\,s on CPU and 550\,ms on GPU.

A greedy baseline was evaluated for comparison. While it achieves faster or similar decision‑making times for small-scale respective medium‑scale scenarios, it consistently results in significantly higher task execution delays, demonstrating that local heuristics cannot replace globally coordinated optimization.

\subsubsection{Communication with Cloud/Edge}
%The evaluation step assesses whether the computation offloading pipeline can accommodate different types of backend servers, namely cloud and edge. As illustrated in Fig.~\ref{fig:plot_requirements} by the colored dots, various servers are selected for offloading based on the expected \glsentrylong{rtt} of a given service and the maximum \glsentrylong{rtt} of a function.

%In addition, the pipeline's ability to handle server failures is evaluated. To simulate a critical, unexpected failure, such as a network disconnect or technical malfunction, the backend servers are shut down sequentially and simultaneously. The tests demonstrate that the pipeline can detect server failures. This enables the \glsentrylong{dm} algorithm to redirect requests to available backend servers or, if none are available, to the local device. However, a delay occurs between the server failure or recovery and its recognition at the Kubernetes control node. In the event of a failure, approximately 30 seconds elapse before the backend server is marked as \textit{not available}, whereas after recovery the server is marked as \textit{ready} immediately, but about 15 seconds are required to restart the pods within the worker node.

% Ersetzt die vorhergenhenden 2 absätze
This evaluation step examines whether the offloading pipeline can accommodate heterogeneous backend servers (cloud and edge). As shown in Fig.\ref{fig:plot_requirements}, the decision‑making mechanism selects different servers depending on the expected RTT of each service and the maximum RTT allowed by the function.

We also assess the pipeline’s robustness to backend failures. To simulate unexpected outages, servers are shut down sequentially and simultaneously. The pipeline reliably detects failures and redirects requests to remaining servers or to local execution if none are available. The observed detection delays stem from Kubernetes’ control‑plane update intervals: a failed server is marked unavailable only after roughly 30\,s, while recovery is detected immediately, but pod restarts require about 15\,s before the server becomes operational again. These delays are specific to Kubernetes, while a lightweight custom API can detect failures within 0.5-2\,s depending on the fault.

\begin{figure}[bt]
    \centering
    \includegraphics[width=\columnwidth]{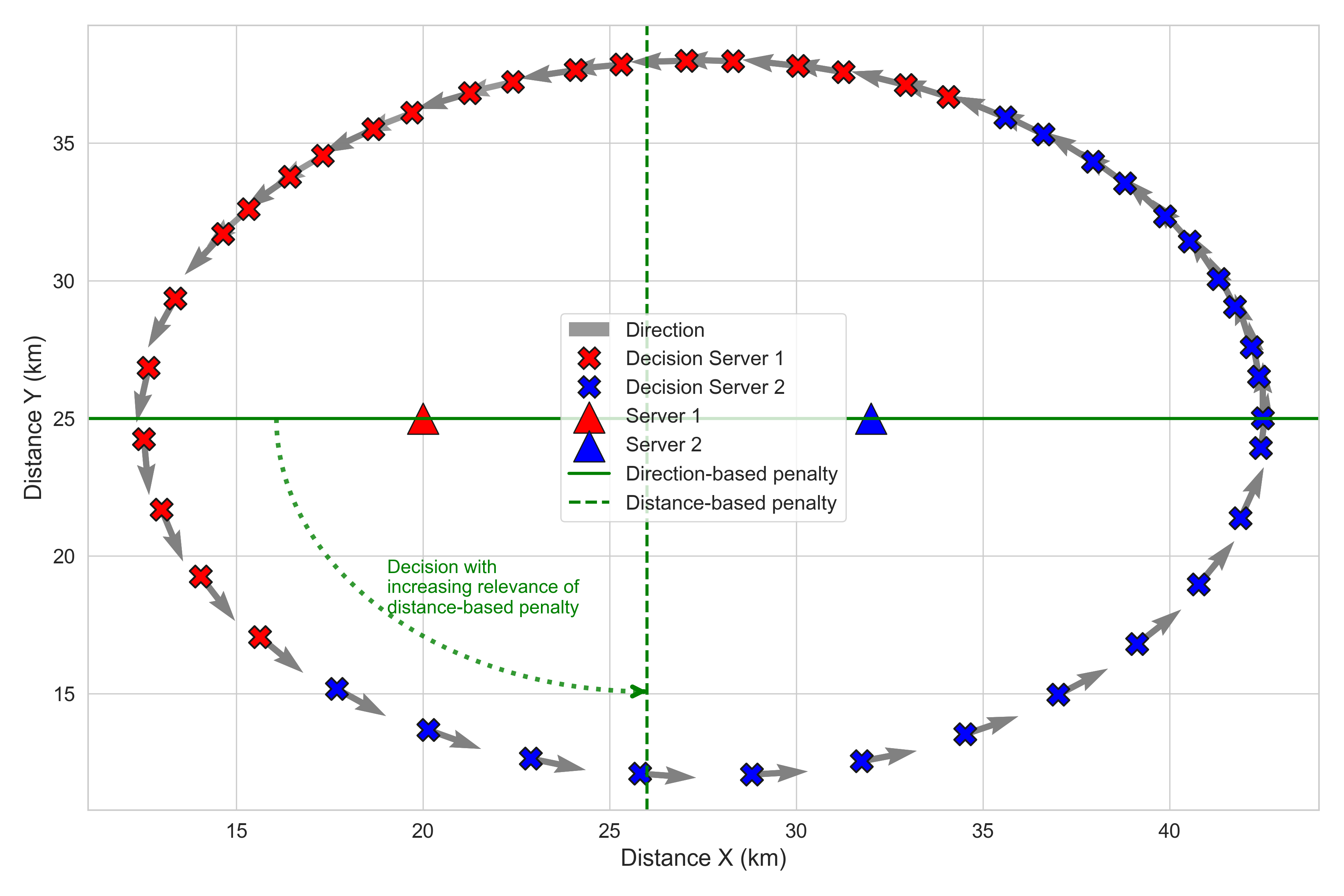}
    \caption{\glsentrylong{pso} decision for a simulated elliptical trajectory around two edge servers, considering direction‑based (62.5\,\%) and distance‑based (37.5\,\%) penalties. Crosses mark the offloading decisions, and colors indicate the selected server. A fixed processing time highlights the influence of both penalties; green lines and arrows illustrate how the decision shifts with changing penalty weights.}
    \label{fig:plot_elliptic}
\end{figure}

\section{Discussion and Limitations}\label{sec:discussion}

The evaluation shows that decisions follow the predefined functional requirements and RTT values sampled from empirically bounded distributions. While this yields correct decisions based on predicted values, measured \gls{rtt}s frequently exceed the thresholds due to network congestion and backend load variability. This exposes the system’s sensitivity to inaccurate runtime assessments: without the Prediction Stage and Detection Layer, the Decision Layer cannot react to \gls{rtt} volatility or accumulated prediction errors. The mismatch between predicted and measured \gls{rtt} therefore highlights the necessity of the full feedback mechanism for real‑time correctness.

Two additional limitations arise in the decision-making process. First, high‑performance servers are consistently preferred because of their lower predicted execution times. Although this reduces latency, it risks overloading these servers when multiple vehicles compete for resources, potentially causing deadline violations for functions that depend on them.

Second, decision‑making time must be considered independently of computation and communication time. The modified \gls{pso} algorithm achieves near‑optimal results with low \gls{cpu} overhead, while the GPU implementation processes 1,000 tasks across 15 server options in roughly 550\,ms. This demonstrates scalability but also emphasizes the need for efficient parallelization to avoid excessive decision‑making delays. 

A greedy baseline was evaluated for comparison. While it achieves faster or similar decision‑making times in small‑ and medium‑scale scenarios and initially produces results close to brute force and \gls{pso}, its performance degrades as the number of tasks and servers increases. For ten or more servers, the greedy method frequently converges to local minima and its computation time scales poorly, resulting in significantly higher task execution delays. This indicates that greedy heuristics are suitable only for small‑scale scenarios and become ineffective as problem complexity grows.

Backend failure handling represents another limitation. While the initial setup required up to 30\,s to mark a failed server as unavailable and an additional 15\,s for recovery, integrating a dedicated failure‑notification API reduced these delays to approximately 2\,s for failure detection and 4\,s for recovery. This improves responsiveness but introduces trade‑offs in monitoring overhead and system stability that must be balanced in future work.

Future work will integrate the Prediction Stage and Detection Layer as active monitoring and learning components, enabling real-time classification of offloading outcomes and adaptive tuning of prediction thresholds to better handle dynamic network conditions and service availability.

\section{Conclusion} \label{sec:conclusion}
%To enable computation offloading in a real-world vehicular context, this paper proposes a scalable and extensible computation offloading pipeline based on a four-layer architecture. The architecture comprises the \textbf{Extraction Layer} which is responsible for providing and managing function- and server-related data, the \textbf{Decision Layer} that implements offloading\glsentrylong{dm} by considering functional requirements, distances to servers, and the vehicle’s direction of movement, the \textbf{Execution and Offloading Layer} to handle the actual execution and delegation of tasks, and the \textbf{Detection Layer} which enables continuous updates based on previous system state classification. The evaluation was carried out using a simulation environment for input data generation, integrated with a cloud instance running a Kubernetes cluster. The main findings can be summarized as follows:
This paper presents a scalable computation offloading pipeline for software‑defined vehicles, structured into four layers: extraction, decision, execution, and detection. Evaluation in a simulated vehicular‑cloud environment shows that the modified \gls{pso} algorithm achieves near‑optimal offloading decisions with an average deviation of 0.7\,\% from brute‑force results while maintaining linear scalability. The system reliably handles heterogeneous backend environments and redirects tasks during server failures, though recovery delays highlight the need for improved failure detection. Key findings include:

\begin{itemize}
    \item By combining functional requirements with direction‑ and distance‑based penalties, the pipeline consistently selects the server with optimal \glsentrylong{rtt} and falls back to local execution when thresholds are exceeded, underscoring the importance of accurate \glsentrylong{rtt} prediction.

    \item In heterogeneous backend environments, the system dynamically manages cloud and edge servers and redirects tasks during failures, demonstrating resilience despite modest detection and recovery delays.

    \item The modified \gls{pso} algorithm achieves near‑optimal accuracy and scales linearly with the number of servers and tasks.
\end{itemize}
 
This study focuses on the pipeline structure and the decision‑making stage. Future work will address the prediction layer—which estimates transmission time, computation time, and utilization—and the detection layer, which enables adaptive feedback based on offloading outcomes. This modular evaluation strategy supports targeted refinement and scalability across pipeline components

\section*{Acknowledgment}
The authors would like to thank the German Federal Ministry of Education and Research (BMBF) (under Grant Number: 16MEE0472) and the Chips Joint Undertaking for the financial support under Grant Agreement No: 101139789 (HAL4SDV). The responsibility for the content of this publication lies with the authors.

The authors disclose that generative AI has been used for improving the grammar and language of the paper. The authors have reviewed and edited all content as needed and take full responsibility for the scientific integrity and authenticity of this article.

\bibliographystyle{IEEEtran}
\bibliography{bib}

\end{document}